\documentclass[12pt]{article}

\usepackage{epsfig}
\usepackage{amsmath}
\usepackage{graphicx}
\usepackage{latexsym}
\usepackage{amsfonts}
\usepackage{url,hyperref}
\usepackage{bm}
\usepackage{textcomp}
\newcommand{\nn}{\nonumber}
\newcommand{\beq}{\begin{equation}}
\newcommand{\eeq}{\end{equation}}
\def\bea{\begin{eqnarray}}
\def\eea{\end{eqnarray}}
\def\bcen{\begin{center}}
\def\ecen{\end{center}}

\begin{document}

\renewcommand{\thefootnote}{\alph{footnote}}


\bcen
{\bf\Large Analysis of the ${\bar B} \to {\bar K}_2(1430) \, \ell^+ \, \ell^-$ decay}
\vskip 1cm
{\bf
S.~R.~Choudhury$^1$\footnote{src@iiserbhopal.ac.in},
A.~S.~Cornell$^2$\footnote{alan.cornell@wits.ac.za}
and Naveen Gaur$^3$\footnote{gaur.nav@gmail.com}
}
\vskip .7cm
{\sl
$^1$CTP, Jamia Millia , New Delhi 110025 and
Indian Institute of Science Education and Research, Bhopal, India,\\
$^2$National Institute for Theoretical Physics; School of Physics, University of the Witwatersrand, Wits 2050, South Africa\\
$^3$Dyal Singh College, University of Delhi, Delhi - 110003, India.
}
\ecen
\vskip 0.5cm
\begin{abstract}

\noindent We present an analysis of the decay process ${\bar B} \to {\bar K}_2(1430)\,  \ell^+ \, \ell^-$, where this process has all the features of the related and well investigated process ${\bar B} \to {\bar K}^*(890) \, \ell^+ \, \ell^-$, with theoretically comparable branching ratios. The differential decay rate as well as the forward-backward asymmetry are worked out, where the sensitivity of these to possible right-handed couplings for the related $b \to s$ radiative decay are also investigated.
\end{abstract}

\renewcommand{\thefootnote}{\arabic{footnote}}
\setcounter{footnote}{0}  


\section{Introduction}\label{sec1}

\par Flavour Changing Neutral Current (FCNC) decays of the $B$-meson are an important tool for investigating the structure of weak interactions and also for revealing the nature of possible physics scenarios beyond the Standard Model (SM). The reason for this is that FCNC decays are forbidden at tree-level in the SM and occur only through higher-order loop graphs. Therefore FCNC processes are very sensitive to possible small corrections that may be a result of any modification to the SM, or from some new interactions. Of the FCNC decays the radiative mode $B \to K^*(890) \, \gamma$ has been experimentally measured, with a lot of theoretical work also having gone into its study. A related decay, $B \to K_2(1430) \, \gamma$ \cite{Aubert:2003zs} has also been observed experimentally, with branching ratios comparable to the well investigated decay $B \to K^*(890) \, \gamma$. The related decay processes with a lepton pair instead of the photon, which have already been seen for the $K^*(890)$ case, can be expected to be seen for the $K_2(1430)$ case, since the branching ratios are comparable. Analysis of this latter process will therefore be a useful complement to the much investigated analysis for the  $K^*(890) \, \ell^+ \, \ell^-$  process for confrontation with theory (such as Forward-Backward (FB) asymmetries), since the analysis probes the effective Hamiltonian in a similar but not identical way. Data on $K_2(1430) \, \ell^+ \, \ell^-$ would thus provide an independent test of the predictions of the SM.

\par In this paper we study the angular distribution of the rare $B$-decay ${\bar B} \to {\bar K}_2(1430) \, \ell^+ \, \ell^-$ using the standard effective Hamiltonian approach and form factors that have already been estimated for the corresponding radiative decay ${\bar B} \to {\bar K}_2(1430) \, \gamma$ \cite{Cheng:2004yj} . The additional form factors for the dileptonic channel are estimated using the Large Energy Effective Theory (LEET) \cite{Charles:1998dr}, which enables one to relate the additional form factors to the form factors of the radiative mode. We note here that the LEET does not take account of collinear gluons and this deficiency is remedied in the Soft Collinear Effective theory (SCET) introduced by Bauer, Fleming, Pirjol and Stewart \cite{Bauer}. However, as they have shown, interactions with collinear gluons preserve the LEET relations between the form factors for a heavy to light decay as long as we ignore terms suppressed by $m/E$, where $m$ is the mass of the light meson and $E$ is its energy in the $B$-meson's rest frame. Our results provide, just as in the case of the $K^*(890)$ resonance, an opportunity for a straightforward comparison of the basic theory with experimental results, which may be expected in the near future for this channel. The paper is organized as follows: In section 2 we will give the relevant effective Hamiltonian and the LEET form factors for the process under consideration. In section 3 we will work out  the expressions for the differential decay rate for the semi-leptonic decay mode under consideration. We will conclude with our results in section 4.


\section{Effective Hamiltonian and Form Factors}\label{sec2}

\par The short distance contribution to the decay ${\bar B} \to {\bar K}_2(1430) \, \ell^+ \, \ell^-$ is governed by the quark level decay $b \to s \, \ell^+ \, \ell^-$, and where the operator basis can be described by the effective Hamiltonian \cite{Ali:2004dq}:
\bea
{\cal H}^{eff} &=& \frac{G_F \alpha}{\sqrt{2} \pi} V_{ts}^* V_{tb} \Bigg[
- 2 C_{7L}  m_b \left(\bar{s}_L i \sigma_{\mu \nu} \frac{q^\nu}{q^2}
  b_R \right) \left(\bar{\ell} \gamma_\mu \ell \right)
- 2 C_{7R} m_b \left(\bar{s}_R i \sigma_{\mu \nu} \frac{q^\nu}{q^2}
  b_L\right) \left(\bar{\ell} \gamma_\mu \ell \right)   \nonumber \\
&& \hspace{2.2cm} + C_9^{eff} \left(\bar{s}_L \gamma_\mu b_L \right)
          \left(\bar{\ell} \gamma^\mu \ell \right) + C_{10} \left(\bar{s}_L \gamma_\mu b_L \right)
          \left(\bar{\ell} \gamma^\mu \gamma_5 \ell \right)
\Bigg] \,\,\, , \label{eq:2}
\eea
where the $C$'s are the Wilson coefficients. $C_{7R}$ in the SM is zero but may arise in models beyond the SM (BSM). As such we will retain this in order to see its effect on some of the experimentally observable quantities. $C_9^{eff}$ includes the short-distance Wilson coefficient as well as long distance effects simulated through the lepton pair being produced by decay of $c {\bar c}$ resonances, where these are fully spelled out in appendix \ref{appendix:c}. The inclusion of a possible $C_{7R}$ in the effective Hamiltonian, which is otherwise absent in the SM, enables one to study the sensitivity of our results BSM. This is similar to the work of  Kim {\it et al.}\cite{Kim:2000dq} for the decay channel $B \to K^*(890) \, \ell^+ \, \ell^-$ channel. Physics BSM often results in non-standard $Z^{\prime}$ coupling to quarks. As far as the effective Hamiltonian is concerned, this results in modifying the values of  $C_9$ and $C_{10}$ away from their SM values. Following, a recent study of such deviations and the constraints imposed on them from known experimental data \cite{Chiang}, we write additive complements to both $C_9$ and $C_{10}$ that we will detail later.

\par We can rewrite the above effective Hamiltonian in the following form:
\bea
{\cal H}^{eff} &=&  \frac{G_F \alpha}{\sqrt{2} \pi} V_{ts}^* V_{tb}
\Bigg[ -  i C_{7L} m_b \frac{q^\nu}{q^2} \left(T_{\mu \nu} + T^5_{\mu \nu}\right)
 \left(\bar{\ell} \gamma^\mu \ell \right)
-  i C_{7R} m_b \frac{q^\nu}{q^2} \left(T_{\mu \nu} - T^5_{\mu \nu}\right)
 \left(\bar{\ell} \gamma^\mu \ell \right)   \nonumber \\
&& \hspace{2.2cm} + {1\over 2}\left( C_9^{eff} - C_{10} \right) (V - A)_\mu
   \left(\bar{\ell}_L \gamma^\mu \ell_L \right) \nonumber \\
&& \hspace{2.3cm} + {1\over2}\left( C_9^{eff} + C_{10} \right) (V - A)_\mu
   \left(\bar{\ell}_R \gamma^\mu \ell_R \right) \Bigg] \,\,\, , \label{eq:3}
\eea
with
\bea
V_\mu &=&  \left(\bar{s} \gamma_\mu b\right) \,\,\, , \label{eq:4} \\
A_\mu &=&  \left(\bar{s} \gamma_\mu \gamma_5 b\right) \,\,\, , \label{eq:5} \\
T_{\mu\nu} &=&  \left( \bar{s} \sigma_{\mu \nu} b \right) \,\,\, , \label{eq:6}  \\
T^5_{\mu\nu} &=&  \left( \bar{s} \sigma_{\mu \nu} \gamma_
  5b \right) \,\,\, . \label{eq:7}
\eea
In equation (\ref{eq:3}) we have used the $(V-A)$ structure for the hadronic part (except for $C_7$). Note that this structure doesn't change under the transformation $V \leftrightarrow - A$ and $T_{\mu \nu} \leftrightarrow T^5_{\mu \nu}$. Furthermore, we can relate the hadronic factors of $T_{\mu\nu}$ and $T^5_{\mu \nu}$ by using the identity\footnote{Where we have used the convention that $\gamma_5= i \gamma^0 \gamma^1 \gamma^2 \gamma^3$ and that $\varepsilon_{0123}=1$.}:
$$
\sigma_{\mu \nu} = - \frac{i}{2} \varepsilon^{\mu \nu \rho \delta}
\sigma_{\rho \delta} \gamma_5 \,\,\, .
$$

\par The hadronic form factors for the ${\bar B} \to {\bar K}_2(1430)$ decay are defined as:
\bea
\langle K_2 (p') | V_\mu  | B (p) \rangle &=& 2  V  \epsilon^{* \alpha \beta} \varepsilon_{\alpha \mu \nu \rho} pÍ^{\nu} p^{\rho} p_{\beta} \,\,\, ,  \label{eq:8} \\
\langle K_2 (p') | A_\mu | B (p) \rangle &=& \epsilon^{* \alpha \beta}  \Big[ 2 A_1 g_{ \alpha\mu} p_{\beta}                                                         +  A_2  p_{\alpha} p_{\beta}p_{\mu}  +  A_3  p_{\alpha} p_{\beta} p'_{\mu} \Big] \,\,\, , \label{eq:9}\\
\langle K_2 (p') |  i   T_{\mu \nu}  q^{\nu} | B (p) \rangle &=&  \frac{2 i U_1}{m_B}  \epsilon^{* \alpha \beta} \varepsilon_{\mu\alpha\lambda\rho} p_{\beta}  p^{\lambda}  p'^{\rho} \,\,\, , \label{eq:10}   \\
\langle K_2 (p') |  i  T^5_{\mu\nu} q^{\nu} | B (p) \rangle &=&  \epsilon^{* \alpha \beta}   \left( \frac{U_2 (p+p')_{\beta} }{m_B} \right)  \Bigg [  g_{\mu\alpha} (p +p'). q  - (p+p')_{\mu} q^{\alpha} \Bigg]  \nonumber \\
&&-  \epsilon^{* \alpha \beta} p_\alpha p_\beta \Bigg[ q_{\mu} - (p+p')_{\mu} \frac{q^2}{(p+p').q}\Bigg] \frac{U_3}{m_B} \,\,\, , \label{eq:11}
\eea
where $\epsilon^{* \alpha \beta}$ is the polarisation vector for the $K_2$.

\par This leads to a matrix element:
\bea
{\cal M} & = & \left( \frac{\alpha G_F \lambda_{CKM}}{2 \sqrt{2} \pi} \right) \epsilon^{* \alpha \beta} \Bigg[ (\bar{\ell} \gamma^\mu \ell) H^V_{\mu \alpha \beta} + ( \bar{\ell} \gamma^\mu \gamma_5 \ell ) H^A_{\mu \alpha \beta} \Bigg] \,\,\, ,
\eea
where
\bea
H^A_{\mu \alpha \beta} & = & C_{10} \Bigg[ 2 V \varepsilon_{\alpha \mu \nu \rho} pÍ^{\nu} p^{\rho} p_{\beta} - 2 A_1 g_{ \alpha\mu} p_{\beta} -  A_2  p_{\alpha} p_{\beta}p_{\mu}  -  A_3  p_{\alpha} p_{\beta} p'_{\mu} \Bigg] \,\,\, , \\
H^V_{\mu \alpha \beta} & = & C_9^{eff} \Bigg[ 2 V \varepsilon_{\alpha \mu \nu \rho} pÍ^{\nu} p^{\rho} p_{\beta} - 2 A_1 g_{ \alpha\mu} p_{\beta} -  A_2  p_{\alpha} p_{\beta}p_{\mu}  -  A_3  p_{\alpha} p_{\beta} p'_{\mu} \Bigg] \nn \\
&& - 2 (C_{7L} + C_{7R}) \frac{m_b}{q^2} \times \left( \frac{2 i U_1}{m_B} \varepsilon_{\mu\alpha\lambda\rho} p_{\beta}  p^{\lambda}  p'^{\rho} \right) \nn \\
&& + 2 (C_{7L} - C_{7R}) \frac{m_b}{q^2} \times \left( \frac{U_2 (p+p')_{\beta} }{m_B} \right)  \Bigg [  g_{\mu\alpha} (p +p'). q  - (p+p')_{\mu} q^{\alpha} \Bigg] \nn \\
&& + 2 (C_{7L} - C_{7R}) \frac{m_b}{q^2} \times \left( p_\alpha p_\beta \Bigg[ q_{\mu} - (p+p')_{\mu} \frac{q^2}{(p+p').q}\Bigg] \frac{U_3}{m_B} \right) \,\,\, .
\eea

\par We  will now relate these form factors using the LEET approach. Using equations (44) -- (48) of J. Charles {\it et al.} \cite{Charles:1998dr} we get:
\bea
V &=& \frac {i  A_1} { m_B  E} \,\,\, , \nonumber \\
V&=& - \frac{i U_1}{m_B^2} \,\,\, , \nonumber \\
U_2 &=& - 2 A_1 \,\,\, \nonumber \\
A_2 &=&0 \,\,\, , \nonumber \\
A_3&=& \frac{2  U_3}{m_B^2} \,\,\, , \label{eq:13}
\eea
where we have taken the limit of the heavy quark mass going to infinity and $E = p . p'/m_B$. Note that with this approach we have introduced no extra hadronic form factors beyond what is required for the radiative mode. Thus, once we are able to describe the radiative mode we have in effect a check on the model from the dileptonic mode. The radiative mode form factors $U_1$, $U_2$ and $U_3$ have been  given by Cheng and Chua \cite{Cheng:2004yj} in their analysis of radiative charmless decays of the $B$-meson using covariant light cone wave functions. We shall use their results.


\section{Kinematics and differential decay rate}\label{sec:3}

\par If we now use the dilepton centre of mass (CM) frame, where $\theta$ shall be the angle between the $K_2$ meson and the $\ell^+$, and $s$ is the energy squared of the outgoing leptons, then we can write our matrix element squared as a function of $\cos\theta$:
\bea
|{\cal M}|^2 &=& A(s) \cos^2 \theta + B(s) \cos\theta + C(s) \,\,\, ,
\eea
where we have defined the $A(s)$, $B(s)$ and $C(s)$ in appendix \ref{appendix:b}.

\par The decay width for the full process can now be expressed as:
\bea
\frac{d\Gamma}{ds d(\cos\theta)} &=& \frac{\lambda^{1/2}(m_B^2 , s , m_K^2)}{2^9 \pi^3 m_B^3} \left( A(s) \cos^2 \theta + B(s) \cos\theta + C(s) \right) \,\,\, ,
\eea
where $\lambda(a,b,c) = a^2 + b^2 + c^2 - 2 a c - 2 a b - 2 b c$ leads to a differential branching ratio of:
\bea
\frac{d\, Br}{ds} & = & \tau_B \frac{\lambda^{1/2}(m_B^2 , s , m_K^2)}{2^9 \pi^3 m_B^3} \left( \frac{2}{3} A(s) + 2 C(s) \right) \,\,\, , \label{eqn:Br}
\eea
where $\tau_{B^0} = (1.525 \pm 0.009)\times 10^{-12} \mathrm{s}$ and $\tau_{B^+}/\tau_{B^0} = 1.071 \pm 0.009$.

\par The normalised FB asymmetry can now be defined as:
\bea
{\mathcal A}_{FB} &=& \frac{ \displaystyle \int_0^1 \frac{d^2 \Gamma}{ds d(\cos\theta)} d(\cos\theta)   - \int_{-1}^0 \frac{d^2 \Gamma}{ds d(\cos\theta)} d(\cos\theta)}{ \displaystyle \int_{-1}^1 \frac{d^2 \Gamma}{ds d(\cos\theta)} d(\cos\theta) } \,\,\, , \label{eq:FBnorm} \\
& = & \frac{3 B(s)}{2 A(s) + 6 C(s)} \,\,\, .
\eea


\section{Results and Conclusion}\label{sec:4}

\par We have followed reference \cite{Kim:2000dq} for the form of the parameterisation of $C_{7L}$ and $C_{7R}$, which automatically takes care of the constraints imposed by experimental data on the radiative decay. Also, it can take into account the possibility that the phase of this term from the SM value ($u=v=0$), although present in the pure radiative decay, would not show up:
\bea
C_{7L} & = & - \sqrt{0.081} \cos x \exp\left( i (u+v)\right) \,\,\, , \nonumber \\
C_{7R} & = & - \sqrt{0.081} \sin x \exp\left( i (u-v)\right) \,\,\, . \label{eqn:c7def}
\eea
To also take into account possible BSM effects on the other SM Wilson coefficients, we write \cite{Chiang}:
\bea
C_9 &=& C_9^{SM} + z \,\,\, , \\
C_{10} &=& -4.546 + y  \,\,\, ,
\eea
where $C_9^{SM}$ is defined in appendix \ref{appendix:c}. $z$ and $y$ above are constrained from radiative and dileptonic decay data by \cite{Chiang}:
\bea
z < - 4.344 \,\,\, ,&& \nonumber \\
y > + 4.669 \,\,\, , && \nonumber \\
1.05 ( z + 4.01 )^2 + 1.05 (y + 4.669)^2 &<& 59.58 \,\,\, , \nonumber\\
0.61 ( z + 3.89 )^2 + 0.61 ( y + 4.669 )^2 &>& 6.38 \,\,\,  .
\eea
This way of parametrizing possible deviations from SM values takes into account the following facts: The decay rates for radiative and dileptonic $K^*(890)$ decay modes of $B$-mesons are in reasonable agreement with SM values. However, the most significant deviation from SM predictions seems to be in  the recent data of the FB asymmetry for the decay $B \to  K^*(890) \, \ell^+ \, \ell^-$ \cite{Wei:2009zv}. Within the context of the effective Hamiltonian, equation (1), such deviations can be accommodated, without significantly disturbing the predictions for the decay rates, by changing the relative signs of the Wilson co-efficients $C_9$ and $C_{10}$ relative to $C_7$. The parametrization above does just that.
	
\begin{table}[t]
\begin{tabular}{|c|c|c|c|c|c|}
\hline
$(x,u,v)$ & $(0,0,0)$ & $(\pi/4, 0,0)$ & $(-\pi/4, 0, 0)$ & $(0.463648, 0, 0)$ & $(\pi/2, 0, 0)$ \\
\hline
$C_{7R}/C_{7L}$ & 0 & - 1 & + 1 & 0.5 & $C_{7L}/C_{7R} = 0$ \\
\hline
$s$(for ${\cal A}_{FB} = 0$) (GeV) & 2.815580528 & 2.087799812 & 1.936986920 & 2.575567261 & 0.1104016881 \\
\hline
\end{tabular}
\caption{\sl Zeroes of the FB asymmetry for selected values of $(x,u,v)$, for $y = z = 0$.}
\label{tab:one}
\end{table}

\begin{figure}[tb]
\begin{center}
\epsfig{file=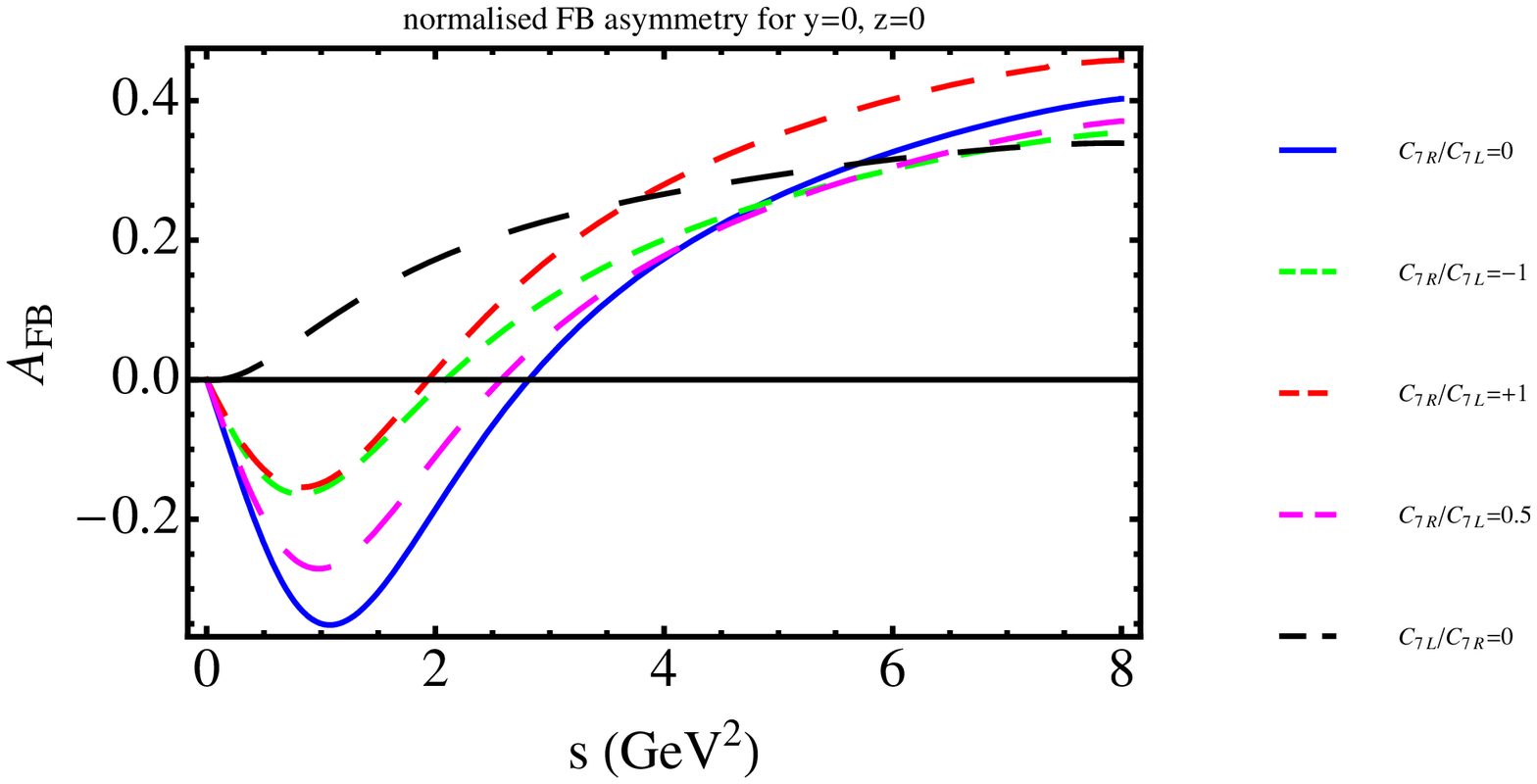,width=.45\textwidth}
\epsfig{file=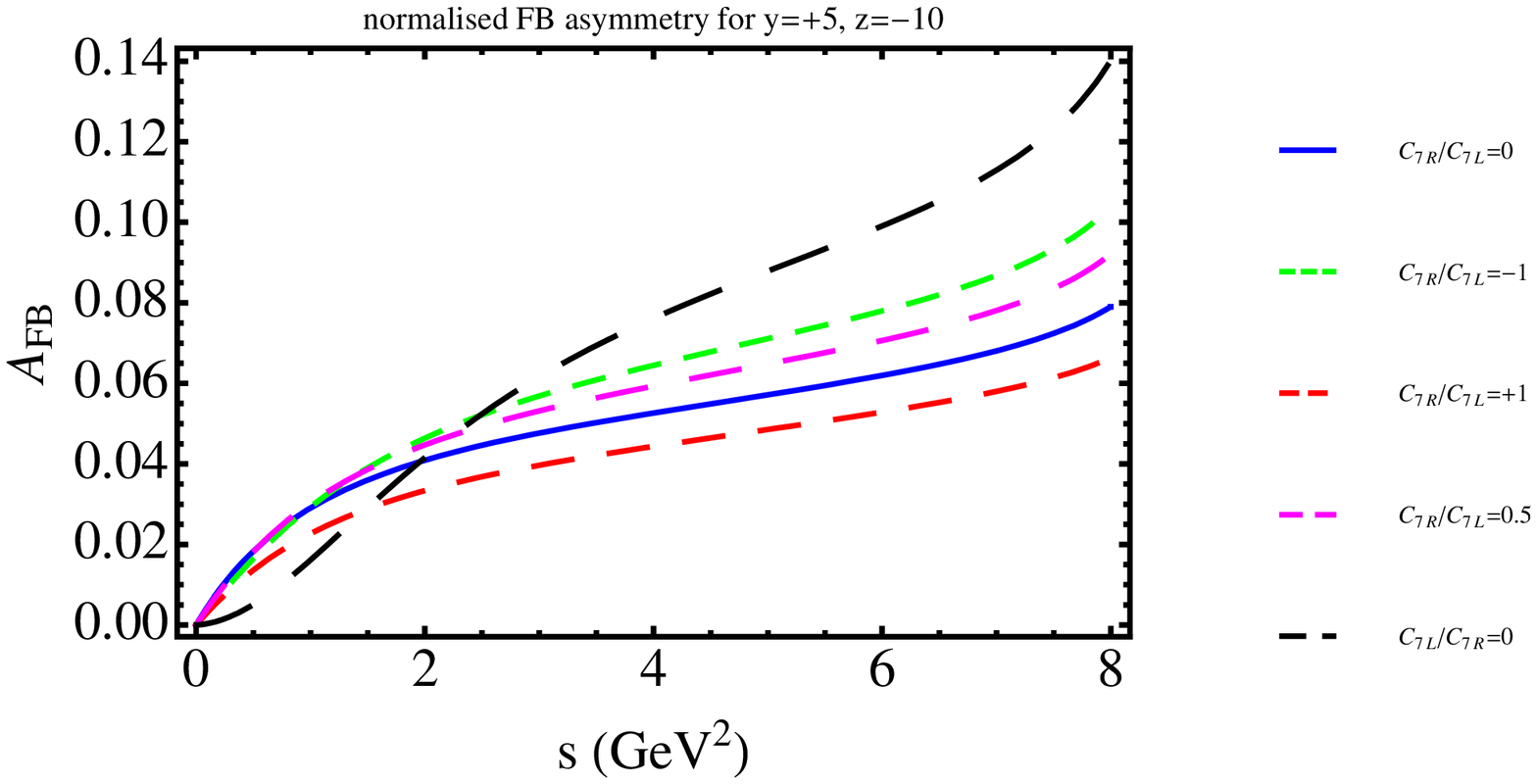,width=.45\textwidth}
\epsfig{file=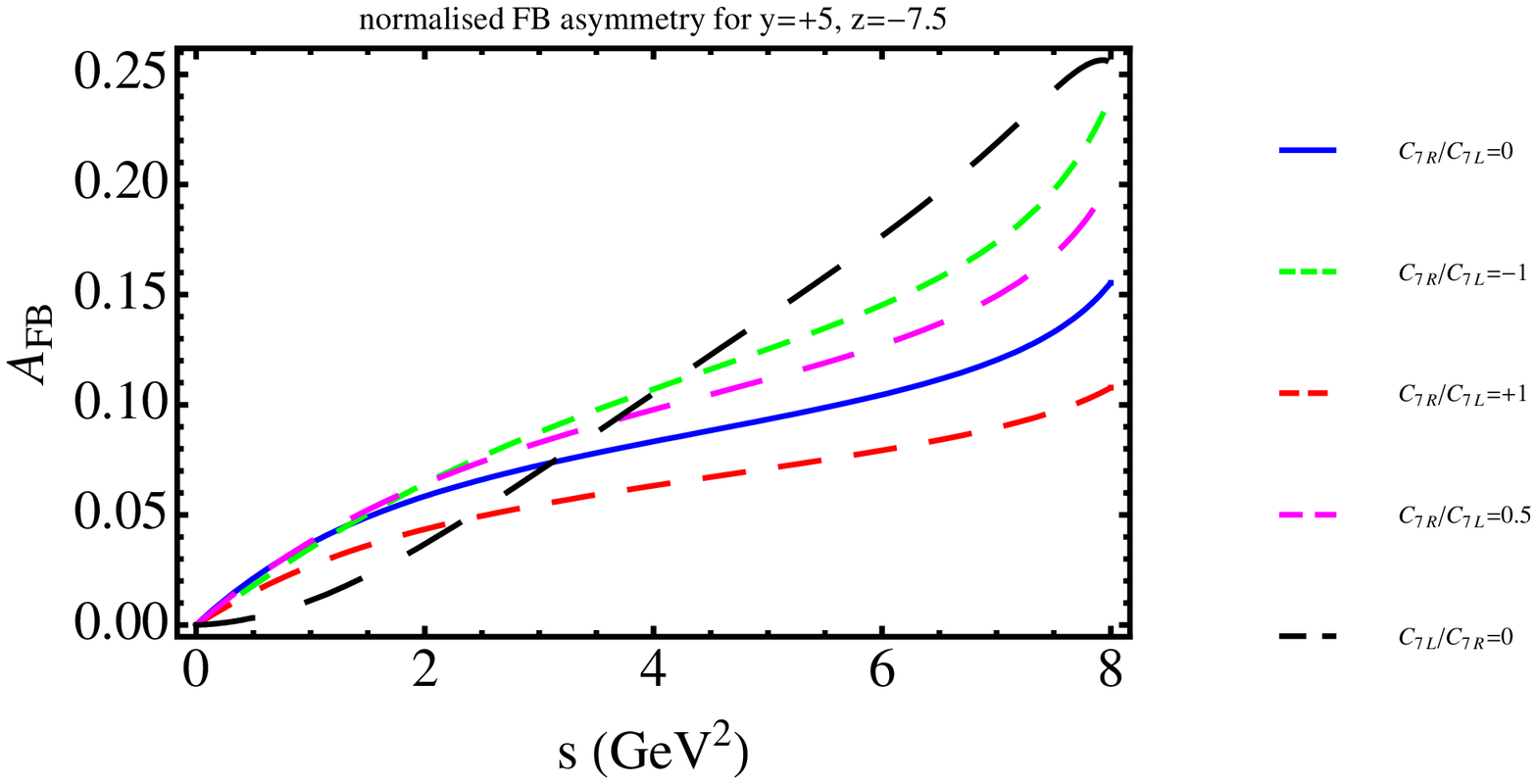,width=.45\textwidth}
\epsfig{file=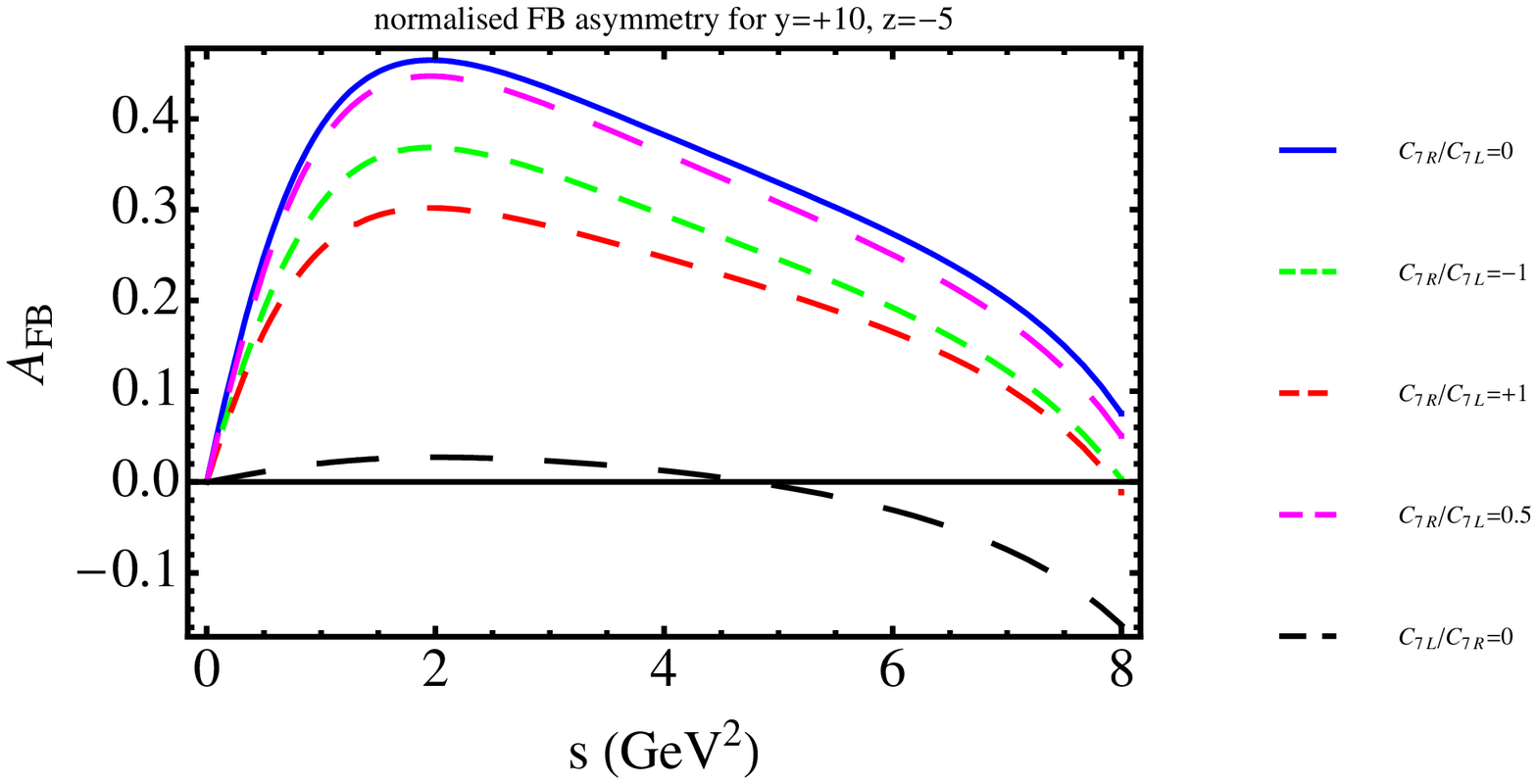,width=.45\textwidth}
\caption{\sl The normalised FB asymmetry (${\cal A}_{FB}$) for a range of values for $C_{7L}$ and $C_{7R}$. The top left panel is for $y = 0$, $z = 0$, the top right panel for $y = +5$, $z = -10$, whilst the bottom left panel is for $y= +5$, $z = -7.5$ and the bottom right panel is for $y = +10$ and $z= -5$. The zeroes of the top left plot are given in table \ref{tab:one}.}
\label{fig:fbnorm}
\end{center}
\end{figure}

\begin{figure}[tb]
\begin{center}
\epsfig{file=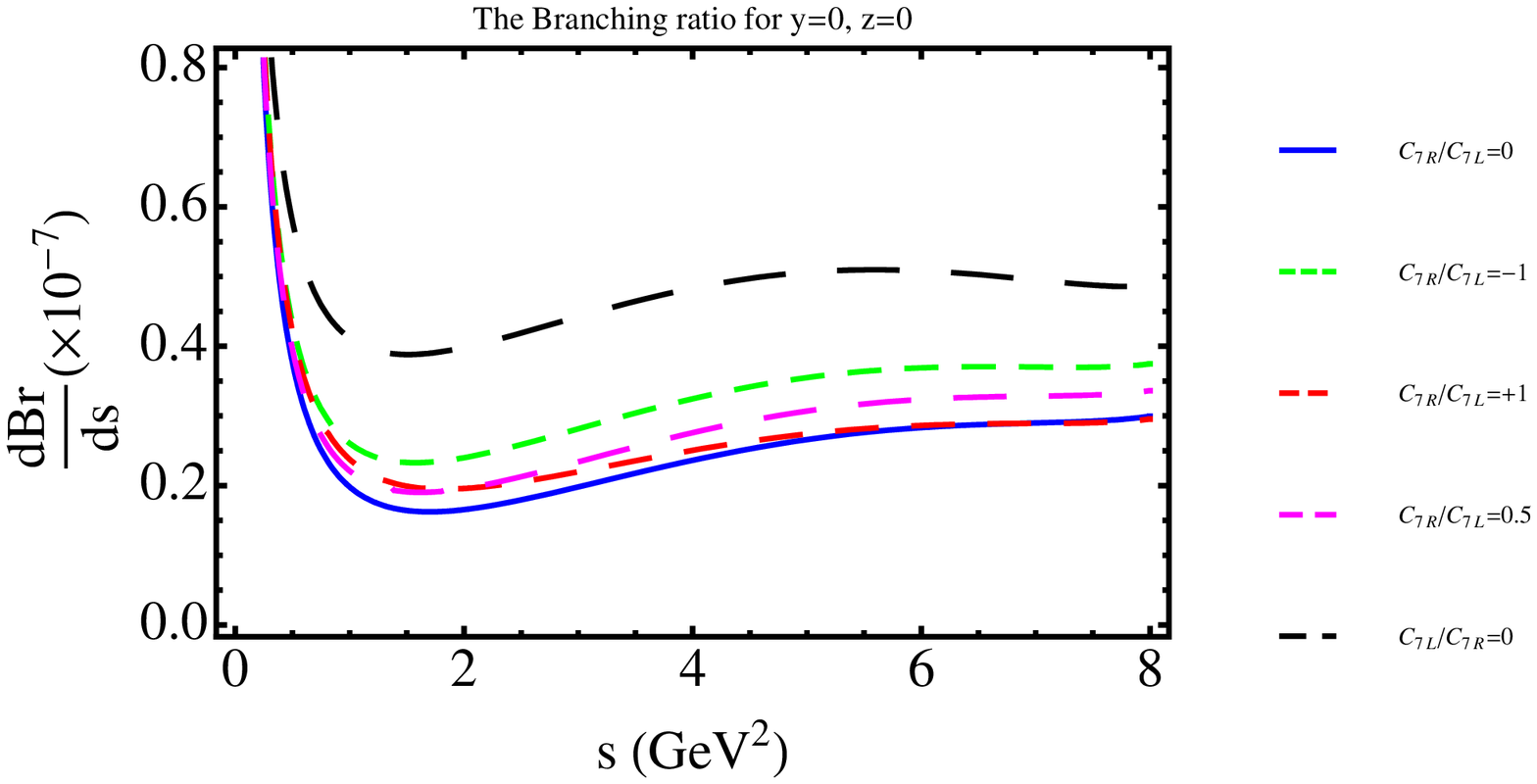,width=.45\textwidth}
\epsfig{file=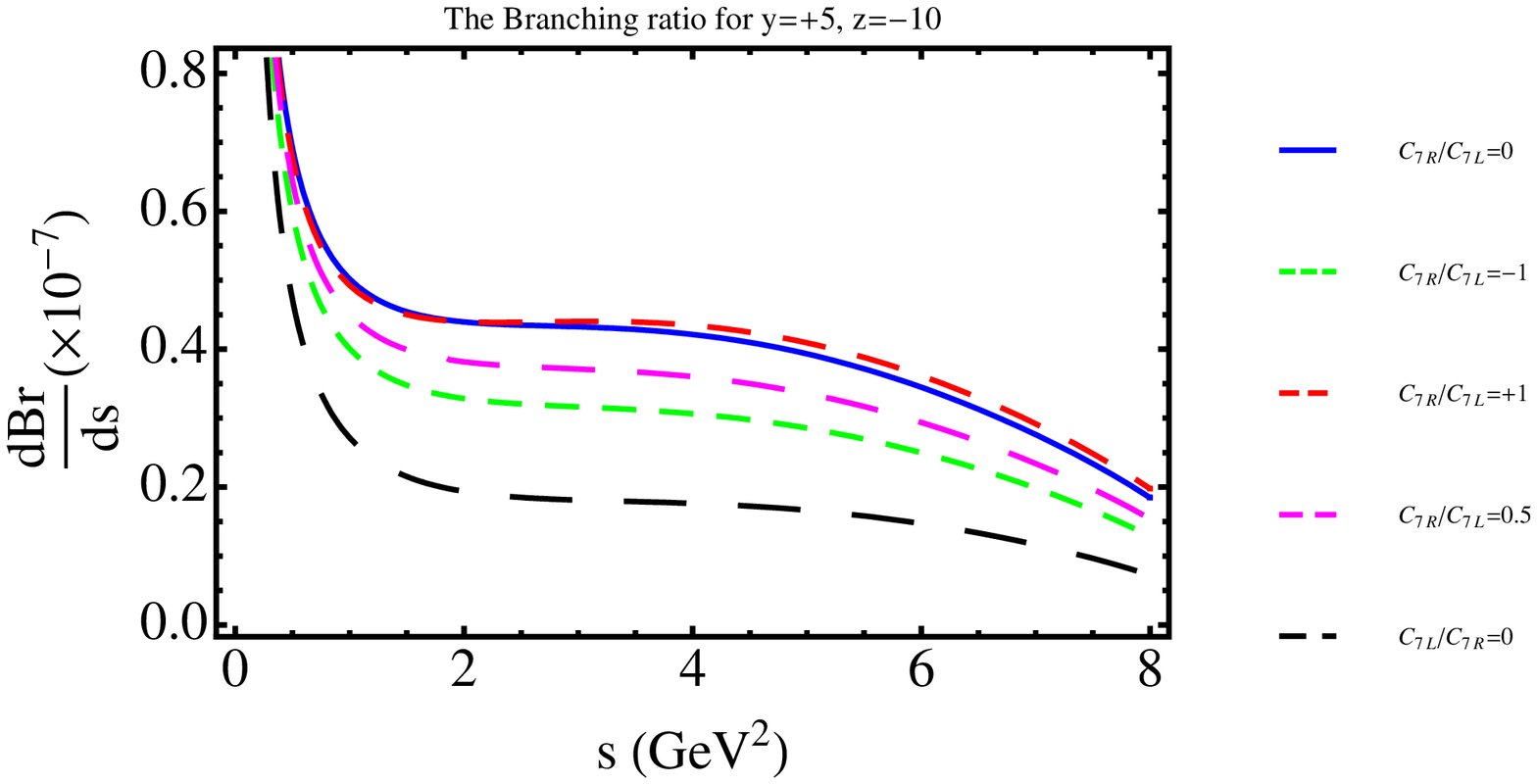,width=.45\textwidth}
\epsfig{file=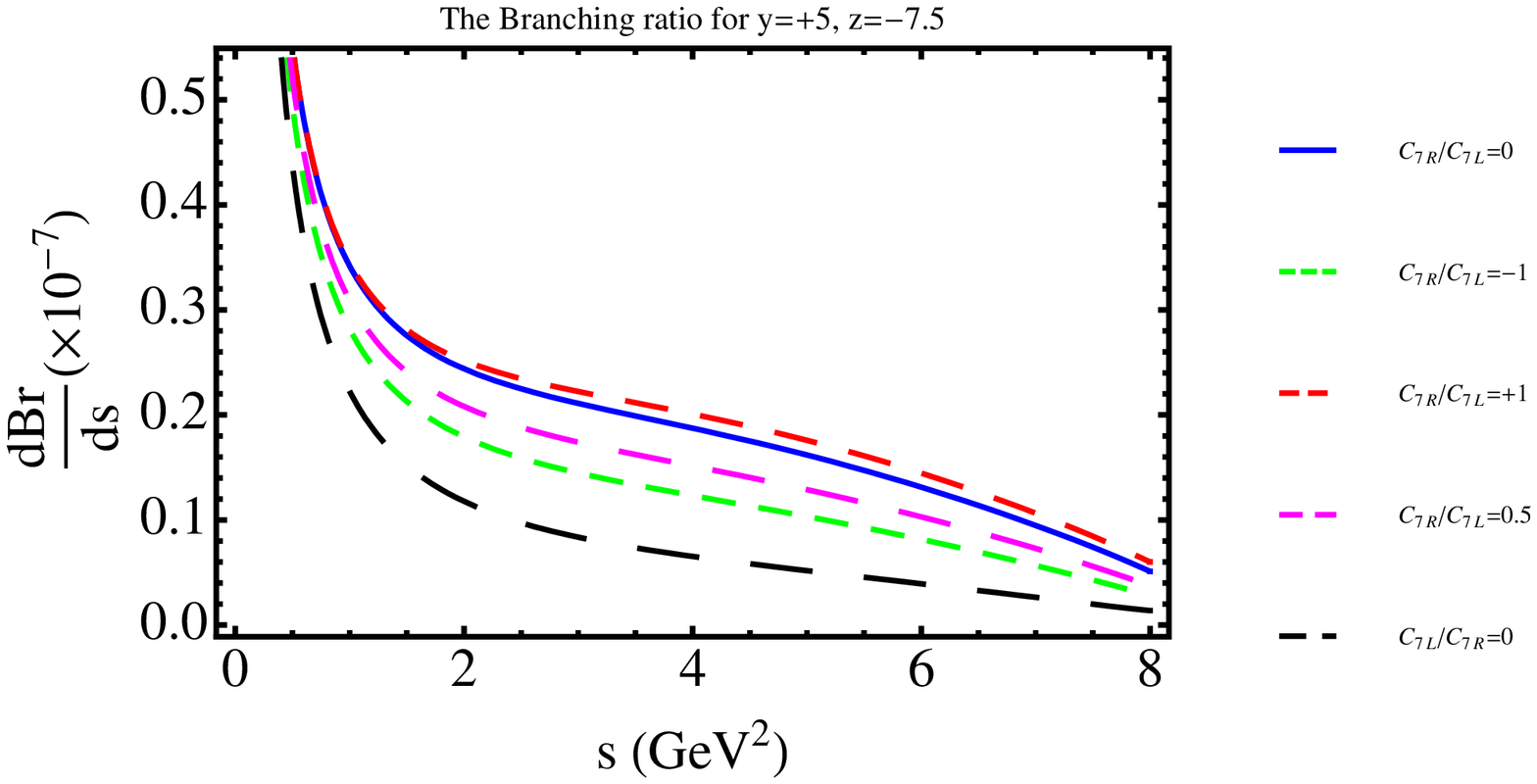,width=.45\textwidth}
\epsfig{file=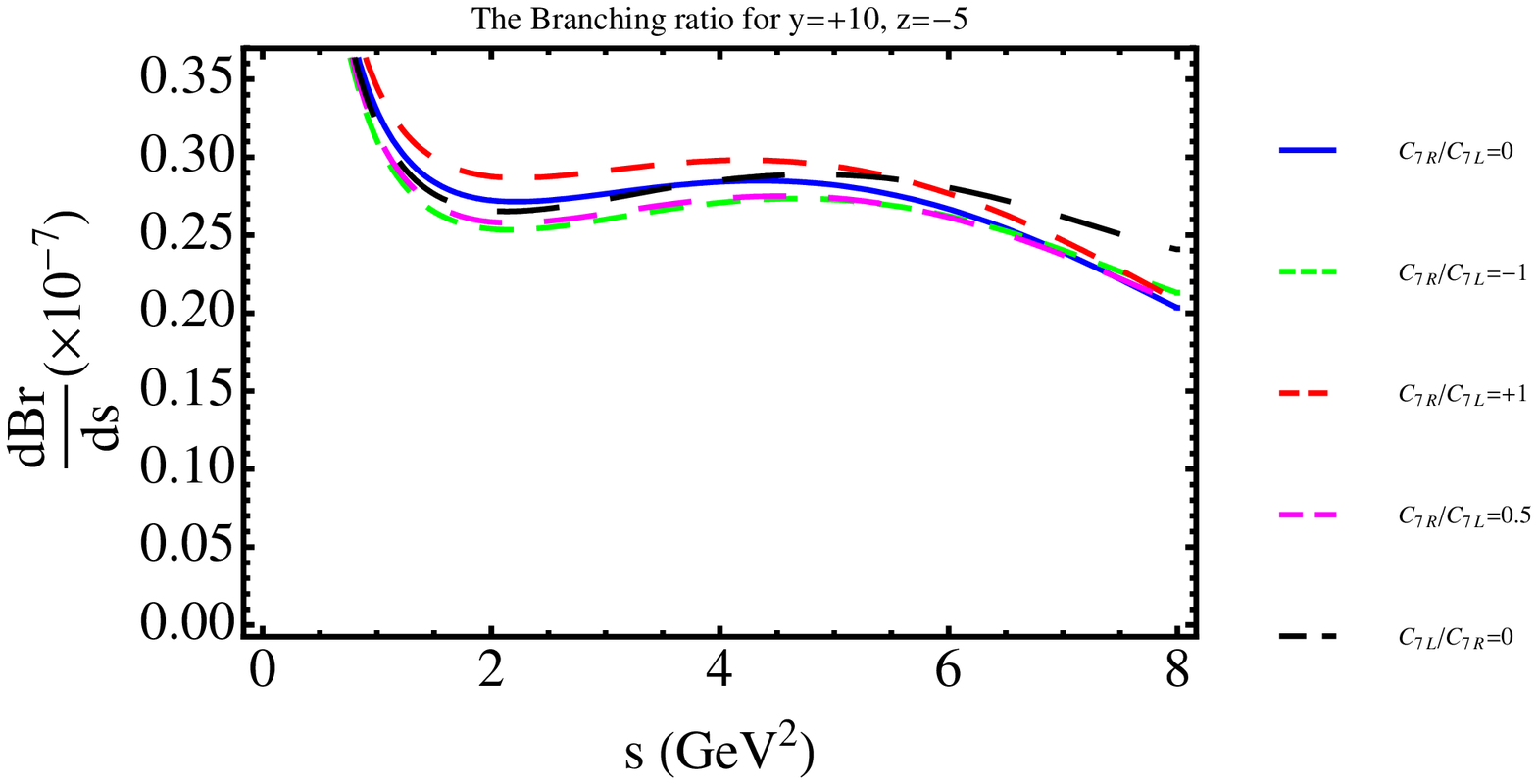,width=.45\textwidth}
\caption{\sl The differential branching ratio, $\displaystyle \frac{d\, Br}{ds}$, as a function of $s$ in the range $0 < s < 8 \, (GeV)^2$.The top left panel is for $y = 0$, $z = 0$, the top right panel for $y = +5$, $z = -10$, whilst the bottom left panel is for $y= +5$, $z = -7.5$ and the bottom right panel is for $y = +10$ and $z= -5$.}
\label{fig:branch}
\end{center}
\end{figure}

\par With this parameterisation we calculate the FB asymmetry and the differential branching ratios (and plot in figures \ref{fig:fbnorm} and \ref{fig:branch}) for some typical values of $(x,u,v)$, as presented in table \ref{tab:one} and these are essentially our results. The results show sensitivity to the values of $x$, $y$ and $z$ quite clearly, where data on the dileptonic decay mode of the ${\bar B} \to {\bar K}_2(1430) \, \ell^+ \, \ell^-$ would be very useful in testing physics BSM . The well measured decay $B \to K^*(890) \, \gamma$ and also $B \to K_2(1430)\, \gamma$ are sensitive only to the sum of the squares of the  magnitudes of $C_{7L}$ and $C_{7R}$,  whereas the decay rate into the dileptonic channel (and especially the zero of the FB asymmetry) change greatly when a right-handed contribution, coming from the $C_{7R}$ term, is present. We have restricted our attention to the large recoil region, where we expect the LEET to be more valid.  The zero of the FB asymmetry, which is a crucial index for comparison with experimental data \cite{Ali:1999mm}, fortunately falls well within this region. We expect that experimental measurements of this dileptonic mode will be available in the near future and the comparison of those with the present theoretical estimates would provide a very useful complement to the corresponding analysis of the well established $K^*(890)$ case. We would also like to note that midway through our calculation, a paper appeared on a similar theme using a somewhat different method of calculation for treating the spin-2 polarization states \cite{Hatanaka:2009gb} leading to the same matrix elements. Our final results are in agreement with their calculations for the SM case, while we also have results for right-handed  $C_7$ couplings and also for possible values of $C_9$  and $C_{10}$ beyond their SM values.

\par Furthermore, just as this work was being completed, a paper from the Belle collaboration \cite{Wei:2009zv} has given indications of new physics BSM, which could be of the type suggested by our equation (\ref{eqn:c7def}) with non-zero phases. Therefore, it would be interesting to also experimentally measure the FB asymmetry in ${\bar B} \to {\bar K}_2(1430)$ transitions, such as we are proposing. For the decay of $B \to K^*(890) \, \ell^+ \, \ell^-$, the Belle result had only 230 events. The number of events expected for the $K_2(1430)$ mode may be estimated by numerical evaluation of equation (\ref{eqn:Br}) and comparing the results with that of the $K^*(890)$ mode. Our estimate gives a non-resonant branching ratio of $4 \times 10^{-7}$. This is in approximate agreement with the result obtained by rough estimation of the area under the curve given as figure 1 of reference \cite{Hatanaka:2009gb}. The corresponding estimate for the $K^*(890)$ mode is approximately $1 \times 10^{-6}$ \cite{Wei:2009zv}. We may thus expect for the $K_2(1430)$ mode roughly half as many events as for the $K^*(890)$ mode. However, the statistics would drastically improve when data from the LHCb becomes available. Estimates made in reference \cite{Rutz} for the LHCb collaboration show that with an integrated annual luminosity of $2 fb^{-1}$, one may expect as many as 8000 $B \to K^*(890) \, \ell^+ \, \ell^-$ events, compared to the 230 events on which the Belle analysis was based \cite{Wei:2009zv}. The number of events for the $K_2(1430)$ mode will correspondingly be of the order of half that many, making analysis of the type suggested in this work experimentally meaningful for comparison with the SM and theories beyond the standard model.


\section*{Acknowledgements}\label{sec:ack}
SRC would like to acknowledge the Department of Science and Technology, Government of India for the awarding of a Ramanna fellowship research grant. NG acknowledges support from the University Grants Commission (UGC), India, under project number 38-58/2009 (SR).


\appendix

\section{The form factors}\label{appendix:a}

\par We shall take the form factors $U_1$, $U_2$ and $U_3$  from Cheng {\it et al.} \cite{Cheng:2004yj}; the remaining form factors can be related to these using the relations in Charles {\it et al.} \cite{Charles:1998dr} as spelt out earlier:
\bea
U_1 (s) & = & \frac{0.19}{1 - 2.22 (s/m_B^2) + 2.13 (s/m_B^2)^2} \,\,\, , \nonumber \\
U_2 (s) & = & \frac{0.19}{\left( 1 - s/m_B^2 \right) \left( 1 - 1.77 (s/m_B^2) + 4.32 (s/m_B^2)^2 \right)} \,\,\, , \nonumber \\
U_3 (s) & = & \frac{0.16}{1 - 2.19 (s/m_B^2) + 1.80 (s/m_B^2)^2} \,\,\, . \nonumber
\eea

\section{The matrix element squared}\label{appendix:b}

\par The matrix element squared in the dilepton CM frame can be written as:
\bea
|{\cal M}|^2 &=& A(s) \cos^2 \theta + B(s) \cos\theta + C(s) \,\,\, ,
\eea
where $A(s)$, $B(s)$ and $C(s)$ are functions involving the form factors and the Wilson coefficients. The expressions for A(s) and C(s), as expected, are very long and we shall not exhibit them. However, the function $B(s)$, which plays the crucial role in the FB asymmetry, is very compact and given below:
\bea
B(s) & = & -\frac{ C_{10} G_F^2\lambda (m_B^2, s, m_K^2 )^{3/2} \alpha^2 \lambda _{CKM}^2  U_1(s) U_2(s)}{2 m_B^3 m_K^2 \pi^2} \nonumber \\
&& \hspace{2cm} \times \left( m_b C_{7R}  m_K^2+ m_b \left(2 m_B^2-m_K^2\right) C_{7L} +m_B s
   \mathrm{Re}(C_9^{eff})(s)\right) \,\,\, .
\eea

\section{Input parameters and Wilson coefficients}\label{appendix:c}

\par The input parameters used in the generation of the numerical results are as follows:
\begin{center}
$m_B = 5.26$GeV ~~,~~ $m_{K^*} = 1.43$GeV ~~,~~ $m_b = 4.8$GeV ~~,~~ $m_c = 1.4$GeV , \\
$m_s = 0.1$GeV ~~,~~ ${\cal B}(J/\psi(1S) \to \ell^+ \ell^-) = 6 \times 10^{-2}$ , \\ $m_{J/\psi(1S)} = 3.097$GeV ~~,~~ ${\cal B}(\psi(2S) \to \ell^+ \ell^-) = 8.3 \times 10^{-3}$ , \\$m_{\psi(2S)} = 3.097$GeV ~~,~~ $\Gamma_{\psi(2S)} = 0.277 \times 10^{-3}$GeV , \\ $\Gamma_{J/\psi(1S)} = 0.088 \times 10^{-3}$GeV ~~,~~ $V_{tb} V_{ts}^* = 0.0385$ ~~,~~ $\alpha = \frac{1}{129}$ ~~,~~ $G_F = 1.17 \times 10^{-5}$ GeV$^{-2}$.
\end{center}

\noindent The Wilson coefficients used were as in Kim {\it et al.} \cite{Kim:2000dq}, namely:
\bea
C_{7L} & = & - \sqrt{0.081} \cos x \exp\left( i (u+v)\right) \,\,\, , \nonumber \\
C_{7R} & = & - \sqrt{0.081} \sin x \exp\left( i (u-v)\right) \,\,\, , \nonumber \\
C_{10} &=& - 4.546 \,\,\, , \nonumber
\eea
\bea
C_9^{SM} & = & 4.153 + 0.381 g\left(\frac{m_c}{m_b} , \frac{s}{m_B^2} \right) + 0.033 g\left(1, \frac{s}{m_B^2} \right) + 0.032 g\left(0, \frac{s}{m_B^2} \right) - 0.381 \times 2.3 \times \frac{3 \pi}{\alpha} \nonumber \\
&& \hspace{1 cm} \times \left( \frac{\Gamma_{\psi(2S)} {\cal B}(\psi(2S) \to \ell^+ \ell^-) m_{\psi(2S)}}{s - m_{\psi(2S)}^2 + i m_{\psi(2S)} \Gamma_{\psi(2S)}} + \frac{\Gamma_{J/\psi(1S)} {\cal B}(J/\psi(1S) \to \ell^+ \ell^-) m_{J/\psi(1S)}}{s - m_{J/\psi(1S)}^2 + i m_{J/\psi(1S)} \Gamma_{J/\psi(1S)}}\right) , \nonumber
\eea
where the function $g$ is taken from reference \cite{gfunc}:
\bea
g(\hat{m}_i, \hat{s}) & = & - \frac{8}{9} \ln (\hat{m}_i) + \frac{8}{27} + \frac{4}{9} \left( \frac{4 \hat{m}_i^2}{\hat{s}} \right) - \frac{2}{9} \left( 2 + \frac{4 \hat{m}_i^2}{\hat{s}} \right) \sqrt{ \left| 1 - \frac{4 \hat{m}_i^2}{\hat{s}} \right|} \nonumber \\
&& \hspace{1 in} \times \left\{ \begin{array}{lc}
\left| \ln \left(\frac{1 + \sqrt{1 - 4 \hat{m}_i^2/\hat{s}}}{1 - \sqrt{1 - 4 \hat{m}_i^2/\hat{s}}} \right) - i \pi \right| & , 4 \hat{m}_i^2 < \hat{s} \\
2 \arctan \frac{1}{\sqrt{4 \hat{m}_i^2/\hat{s} - 1}} & , 4 \hat{m}_i^2 > \hat{s}
\end{array} \right. \,\,\, . \nonumber
\eea


\end{document}